\documentclass[sigconf, anonymous=false]{acmart}

\usepackage{array}
\AtBeginDocument{%
  \providecommand\BibTeX{{%
    \normalfont B\kern-0.5em{\scshape i\kern-0.25em b}\kern-0.8em\TeX}}}

\setcopyright{acmlicensed}
\copyrightyear{2024}
\acmYear{2024}
\acmDOI{XXXXXXX.XXXXXXX}

\acmConference[SIGIR '24]{In Proceedings of
the 47th International ACM SIGIR Conference on Research and Development
in Information Retrieval}{July 14--18,
  2024}{Washington, DC}
\acmISBN{978-1-4503-XXXX-X/18/06}

\begin{document}

\title{Enhancing Question Answering Precision with Optimized Vector Retrieval and Instructions}


%% \author{Anonymous}

\author{Lixiao Yang}
\affiliation{%
  \institution{Drexel University}
  \streetaddress{3141 Chestnut St}
  \city{Philadelphia}
  \state{PA}
  \postcode{19104}
  \country{USA}}
\email{ly364@drexel.edu}

\author{Mengyang Xu}
\affiliation{%
  \institution{Drexel University}
  \streetaddress{3141 Chestnut St}
  \city{Philadelphia}
  \state{PA}
  \postcode{19104}
  \country{USA}}
\email{mx73@drexel.edu}

\author{Weimao Ke}
\affiliation{%
  \institution{Drexel University}
  \streetaddress{3141 Chestnut St}
  \city{Philadelphia}
  \state{PA}
  \postcode{19104}
  \country{USA}}
\email{wk@drexel.edu}

\renewcommand{\shortauthors}{Yang, Xu, and Ke}
% \renewcommand{\shortauthors}{Anonymous}

\begin{abstract}
% TODO: Summarize the main goal, method, and findings of the paper (100-150 words).
Question-answering (QA) is an important application of Information Retrieval (IR) and language models, and the latest trend is toward pre-trained large neural networks with embedding parameters. Augmenting QA performances with these LLMs requires intensive computational resources for fine-tuning. We propose an innovative approach to improve QA task performances by integrating optimized vector retrievals and instruction methodologies. Based on retrieval augmentation, the process involves document embedding, vector retrieval, and context construction for optimal QA results. We experiment with different combinations of text segmentation techniques and similarity functions, and analyze their impacts on QA performances. Results show that the model with a small chunk size of 100 without any overlap of the chunks achieves the best result and outperforms the models based on semantic segmentation using sentences. We discuss related QA examples and offer insight into how model performances are improved within the two-stage framework.  
\end{abstract}






%% \begin{teaserfigure}
%%  \includegraphics[width=\textwidth]{sampleteaser}
%%  \caption{Seattle Mariners at Spring Training, 2010.}
%%  \Description{Enjoying the baseball game from the third-base
%%  seats. Ichiro Suzuki preparing to bat.}

\received{8 February 2024}
%%\received[revised]{12 March 2024}
%%\received[accepted]{5 June 2024}

\maketitle

\section{Introduction}
Accurately answering queries utilizing large-scale datasets has become an important task in the fast evolving filed of natural language processing (NLP). Recent advancements in machine learning and artificial intelligence have led to the development of sophisticated models that can parse, follow, and respond to natural language queries with unprecedented accuracy. Among these, the Retrieval-Augmented Generation (RAG) architecture represents a significant leap forward, combining the strengths of information (vector) retrieval and generative models for QA applications\cite{10.1162/tacl_a_00530}. This paper introduces a novel methodology that leverages the RAG architecture, aiming to enhance QA precision through optimized vector retrieval and improved context construction. 
% TODO: Introduce the problem area, significance of the research, and a brief overview of GPT+IR.
% TODO: State the paper's main contribution.

\section{Background}

% TODO: Discuss existing work related to GPT, information retrieval, and their combination.
% TODO: Introduce the concepts of GPT, IR, and why their integration is beneficial.

\subsection{IR for Question Answering}
% TODO: 
% Discuss traditional information retrieval and retrieval-based approaches to question answering tasks
Information Retrieval (IR) is pivotal in processing and managing exponentially growing volumes of big data and has played an important role in supporting QA tasks. Traditionally, information retrieval has focused on extracting relevant information from large datasets or corpus, which has become increasingly complex with the exponential growth of digital data. With advances in computing power, IR techniques have evolved from simple keyword-based approaches to sophisticated algorithms utilizing neural networks and deep learning. 

For QA tasks based on a specific corpus, retrieval-based systems have been a popular approach. These systems retrieve the answer to a query by searching a database of indexed information or utilizing various methods like bag-of-words representation including various retrieval functions, morpho-syntactic analysis including stemming and lemmatization, and semantic classification including both deterministic and probabilistic approaches to ensure the efficiency and accuracy of IR results.\cite{KOLOMIYETS20115412} However, these systems often face challenges such as ensuring the accuracy and relevance of the retrieved information, especially in complex query environments requiring additional information contexts. %Based on the RAG architecture, we build a process for question-answering and aim to address these challenges by integrating optimized embedding, retrieval, and ultimately answer generation. We study various text structure sizes and how their combinations impact answer predictions. The objective of this research is improving the performance of QA systems by leveraging the advantage of IR in identifying relevant information for the necessary context and the effectiveness of pre-trained large language models in generating natural language answers in the provided context.

\subsection{NLP and Language Models}
% TODO: 
% Discuss traditional NLP approaches and language models
% Applications for information retrieval, summarization, and question answering

The Transformer architecture proposed by Vaswani et al. \cite{DBLP:journals/corr/VaswaniSPUJGKP17} represents a major innovation in the field of language modeling, particularly relevant to many NLP tasks including language translation, question-answering, and summarization. The modeling architecture relies on the attention mechanism to map global dependencies in the sequences of inputs and outputs. Models such as GPT \cite{radford2018improving} and BERT \cite{DBLP:journals/corr/abs-1810-04805} are built on top of this architecture, demonstrating enhanced language comprehension and generation capabilities. The Transformer architecture provides a solid foundation for research to improve QA accuracy by incorporating pre-trained GPT4ALL and Llama models with the LangChain framework. The Transformer architecture enhances the performance of QA systems by advancing retrieval techniques and instruction methods, particularly in RAG, demonstrating its remarkable ability to process and understand natural language.

\subsection{Retrieval Augmentation and LMs}

% Discuss approaches to integrate information retrieval (with enriched vectors) for language modeling
% Specifically, two lines of research: REALM (or in context REALM), and RAG. 
% TODO for Mengyang: One paragraph of research in REALM or or in context REALM with citations.
Related research on REALM \cite{pmlr-v119-guu20a} and in-context REALM \cite{10.1162/tacl_a_00605} employs a retrieval process to select relevant contexts from a large corpus based on the input question and to generate answers based on that retrieved context. These systems enrich documents and queries (questions) with embedding to facilitate matching. %This used the concept based on in-context REALM, which also relies on embeddings to perform the retrieval step effectively. The embeddings allow the model to understand the semantic similarity between the query and the documents, improving the retrieval's relevance.

% TODO for Lixiao: One paragraph of research in RAG with proper citations. 
The Retrieval Augmented Generation (RAG) architecture emerges as a pivotal innovation, particularly for its integration of Information Retrieval (IR) and Large Language Models (LLMs). %RAG fundamentally enriches these models by integrating extensive knowledge bases, thereby enabling them to handle intricate queries and produce nuanced, contextually rich responses. 
RAG leverages pre-trained models with pre-existing, broad knowledge sources instead of relying solely on model retraining with domain-specific data. This approach has been proven not only enhances the depth and diversity of the model's responses but also substantially reduces training costs and resource requirements, a notable advantage over conventional fine-tuning techniques\cite{pmlr-v119-guu20a}. 

In the context of our research, RAG's application aligns seamlessly with our two-stage QA framework, based on which we can improve the relevance of generated answers, particularly in QA application where the local context is critical. This framework streamlines the overall QA process to deliver context-accurate answers.

% TODO
% Discuss potential gaps and what to be done/improved in future research. 

\section{Methodology}
% TODO: Describe the GPT+IR system architecture.

\subsection{RAG Architecture}

Our approach is based on the RAG architecture that combines the strengths of both retrieval-based and generative models for question answering. The workflow can be described with the following major processes: 

% TODO: Explain how the data was prepared, including any splitting and chunking strategies used.
{\textbf 1. Document Pre-Processing:}
\begin{enumerate}
    \item Initial Preparation: This involves cleaning the text, removing formatting and irrelevant information, and segmenting the text into manageable chunks, e.g. of fixed sizes or structures of sentences or paragraphs.
    \item Chunking: The pre-processed documents are divided into smaller parts, such as paragraphs, sentences, or custom-sized chunks. This step is crucial for efficient retrieval and ensuring relevant context is available for the generative model.
\end{enumerate}

{\textbf 2. Vector Embedding}
\begin{enumerate}
    \item Document Embedding: Each chunk of text is transformed into a vector representation using embedding techniques. This involves using pre-trained models like BERT or other language models.
    \item Embedding Storage: The vectors are then stored in a database or a vector store, which allows for efficient similarity search.
\end{enumerate}

{\textbf 3. Query Embedding and Matching/Retrieval}
\begin{enumerate}
    \item Query Processing: When a question is asked, it is also enriched using a similar embedding process.
    \item Vector Matching: The query vector is then matched against the document vectors in the vector store. This involves calculating similarity scores and retrieving the most relevant chunks based on these scores.
\end{enumerate}

{\textbf 4. Answer Generation}
\begin{enumerate}
    \item Context Formation: The retrieved chunks form the context is then fed to a generative language model.
    \item Answer Synthesis: The language model, often a large transformer-based model like GPT, generates an answer based on the provided context. This step involves understanding the question, processing the context, and synthesizing a coherent and accurate response.
\end{enumerate}

\subsection{Two-Stage Question Answering}

In this research, we focus on two stages affecting answer accuracy. The first stage in getting to the correct answer is to identify and retrieve relevant information (chunks) that either contains the answer itself or provides sufficient context for the answer to come out. Several techniques associated with text chunking, overlapping, embedding, and matching (ranking) are crucial for this vector-retrieval stage. 

The second stage, once the context can be constructed with retrieved chunks, is to adopt a pre-trained generative model and provide proper instructions with the context to identify the answer. For this study, we use a small model, namely Mistral 7B Instruct\cite{jiang2023mistral}, fine-tuned to follow instructions and focus on providing the clear, specific instructions with the context to help it produce precise answers. %For datasets focused on answer precision, there are combinations of model selection and instruction construction (prompt engineering) that potentiate optimal outcomes. 
% Assuming sufficient retrieved context/context for the answer, crafting effective prompts is crucial. Prompts guide the model's focus and can significantly impact the quality of the answer. Clear and specific instructions--within a well-constructed template with the original question, context, and examples--can lead to better-targeted answers. 

%Here are factors we consider in the experiments: 

% \begin{itemize}
%     \item Model Type: The choice of generative model (e.g., LLaMa models) affects the quality of the generated answers. Models trained specifically for question-answering or with domain-specific knowledge can be more effective. In this study, we use the Mistral
%     \item Prompt Engineering: Assuming sufficient retrieved context/context for the answer, crafting effective prompts is crucial. Prompts guide the model's focus and can significantly impact the quality of the answer. Clear and specific instructions--within a well-constructed template with the original question, context, and examples--can lead to better-targeted answers. 
% \end{itemize}

% TODO: Explain the data sources, including the Stanford QA dataset.
\subsection{Benchmark Datasets}
% TODO: Describe the experimental setup, including the datasets used (NewsQA, QAConv).

We conduct our experiments on the following datasets: 
\begin{itemize}
    \item NewsQA Dataset: This dataset comprises a collection of 12,744 CNN news from the DeepMind Q\&A dataset accompanying with over 119,633 paired questions. The answers are spans of text from the corresponding articles, necessitating reasoning and comprehension beyond simple word matching. This dataset is particularly relevant for its characteristics of real-world, high-volume, and broad range of topics, making it an ideal testbed for our research. %This is developed by Microsoft Research Montreal \cite{DBLP:journals/corr/TrischlerWYHSBS16}.
    \item QAConv Dataset: This dataset is a novel dataset for question answering(QA) that leverages conversations as the primary source of information. Our emphasis is on detailed conversations, such as those found in business emails, panel discussions and professional channels, which are characteristically lengthy, intricate, asynchronous, and rich in domain-specific knowledge. The dataset comprises 34,204 QA pairs, which include various types of questions like span-based, free-form, and those without answers. These were derived from 10,529 carefully chosen conversations, incorporating questions generated by both humans and machines. %This is developed by Kelvin Guu and his team \cite{wu2021qaconv}%. TODO for Mengyang: Add more details about the dataset and include some basic statistics. 
\end{itemize}

These datasets have been used in related question-answering research and are focused on identifying precise event-related answers in news reports or conversations.

% TODO for Lixiao: write about details of specific models used in the pipeline, and different chunking strategies, and retrieval methods.
\subsection{Variables}

\textbf{Models.} Our research on optimizing QA systems integrates Mistral 7B, a model with 7.3 billion parameters, due to its superior performance over Llama 2 13B and comparable results with Llama 1 34B, especially in commonsense reasoning and reading comprehension. Mistral 7B's Grouped-query attention and Sliding Window Attention features are pivotal for efficiently processing lengthy text sequences, ensuring contextual accuracy in complex QA scenarios\cite{jiang2023mistral}. %This model's adaptability and effectiveness in handling diverse queries from specialized domains significantly enhance our system's accuracy and reliability in QA tasks. 
% In the context of our research focused on optimizing QA systems through vector retrieval techniques, the incorporation of Mistral 7B, a model with 7.3 billion parameters, is critical. This model demonstrates its excellence by outperforming Llama 2 13B across various benchmarks and showing comparable outcome to Llama 1 34B, particularly in areas requiring commonsense reasoning, world knowledge, and advanced reading comprehension. The unique features of Mistral 7B, such as Grouped-query attention (GQA) and Sliding Window Attention (SWA), enhance its capability to efficiently process extended text sequences, a critical factor for maintaining contextual accuracy in complex QA tasks\cite{jiang2023mistral}. The adaptability of Mistral 7B, as evidenced by its successful fine-tuning for chat-based applications, aligns seamlessly with our experimental requirements, especially in handling diverse and intricate queries drawn from specialized domains like news and conversational datasets. By integrating Mistral 7B into our system, we leverage its computational efficiency and advanced processing capabilities to improve the accuracy and reliability of the QA process.

\textbf{Chunking Strategies.} Diverse chunking strategies are applied to enhance the efficiency of IR and ensure the availability of pertinent context for the generative model. % The segmentation of pre-processed documents into smaller units such as paragraphs, sentences, or custom-sized chunks is a strategic approach that addresses various aspects of query processing. 
\begin{enumerate}
    % \item Paragraph-level chunking is adept at retaining broader context, crucial for queries that necessitate a comprehensive understanding of the subject matter.
    \item Sentences are important semantic structures and provide the baseline chunking strategy on extracting specific information in them.
    \item Chunks and overlap: The adoption of custom-sized chunks combined with overlaps between chunks allowing the system to retrieve different combinations of data pieces for context construction. 
\end{enumerate}

% This multi-faceted chunking methodology ensures that the retrieved information is both relevant and optimally connected for the subsequent generative model, thereby enhancing the overall effectiveness of the QA system.

%% Note: reorganize the specific function names?
\textbf{Retrieval Methods.} The retrieval methods in our QA structure rely on the advanced use of HuggingFaceInstructEmbeddings for embedding both story texts and questions. Texts are segmented into sentence-level or custom-sized chunks utilizing SemanticChunker, each embedded to form dense vector representations stored in a Chroma vectorstore. Retrieval methods mainly include two key distance functions: pairwise and cosine similarity. While pairwise distance is sensitive to vector magnitude, cosine similarity is a normalized score and not influenced by the size of the segment (e.g. sentences of different lengths). % These functions enable the system to select the most contextually relevant text segments based on semantic similarity scores. The combination of strategic text segmentation along with embedding techniques forms the core of our retrieval process.

\subsection{Software and Hardware Setup}

% TODO: Detail the implementation process, including exploration with LangChain and GPT4ALL.

The implementation of the RAG architecture for question answering in these experiments utilizes two key platforms: LangChain and PyTorch. LangChain serves a pivotal role in orchestrating the flow of data through the various stages of the process. It acts as a bridge, integrating the language models used for embedding and retrieval with the generative models responsible for answer generation. % This platform streamlines the interaction between the retrieval systems and the language models, supporting efficient processing of text chunks, document embeddings, and the subsequent context delivery to the generative models.  
PyTorch, on the other hand, is employed primarily for the development and operation of neural networks. It provides the computational backbone for embedding models, for the initial transformation of text data into vector representations, and handling complex tensor operations with GPU acceleration. We conduct the experiments on a MacStudio equipped with an M2 Max chip, which includes 64 GB of memory, 2 TB storage, and 37 GPU cores optimized for Metal Performance Shaders (MPS) acceleration. % This configuration provides a robust platform for the computationally demanding tasks while simulating a real scenario where such models can be deployed in personal computing settings.

\subsection{Evaluation Metrics}
% TODO: Explain the criteria for performance evaluation (e.g., F1 score, exact match).
We perform stemming on the words in the ground truth and predicted answers before using the following metrics to evaluate results:
\begin{itemize}
    \item Exact Match (EM): Measures whether the model's answer exactly matches the ground truth answer.
    \item Precision: Measures the fraction of words in the predicted (generated) answer that match what's in the ground truth. 
    \item Recall: Measures the fraction of words in the ground truth that appear in the generated answer. 
    \item F1 Score: Harmonic mean of precision and recall, accounting for partial overlaps between the predicted answer and the ground truth.
    % \item Response Time: Measures the efficiency of the model in retrieving and generating answers, important for practical applications in personal computing environments.
\end{itemize}

\section{Experiments}

\subsection{Examples of Optimization}
% \begin{comment}
% Best from the baselines compared to (improved) results from models with 1) the Mistral Instruct model that is responsive to instructions, and 2) improved instructions for better answer precision.
% \end{comment}

Base on results from initial experiments, several optimizations are implemented to address the challenges in retrieval strategies, prompt structure, and the choice of LLMs. These changes aim at enhancing the overall effectiveness and efficiency of the question-answering system. We offer examples in this section to showcase the improvement via our two-stage framework. 

\textbf{Stage 1: Retrieval Strategies.} Initial observations indicate that inaccuracies in answers are often due to incorrect retrieved sentences. For instance:\\
\fbox{%
\begin{minipage}{233pt}
\texttt{%
    \textbf{Question:} What was Moninder Singh Pandher acquitted for? \\
    \textbf{Retrieved Sentence:} The Allahabad high court has acquitted Moninder Singh Pandher, his lawyer Sikandar B. Kochar told CNN. \\
    \textbf{Predicted Answer:} Moninder Singh Pandher \\
    \textbf{Actual Answer:} the killing of a teen
}
\end{minipage}
}

To rectify this, we shift to a chunk-based retrieval strategy. The story text is divided into various chunks, and for each retrieved sentence, related chunks are back-retrieved to form the search space for the QA process. This approach provides LLMs with a more comprehensive context, enhancing the accuracy of the answers. Additionally, different chunk sizes and overlapping percentages are experimented with to investigate their impact on retrieval accuracy and efficiency.

\textbf{Stage 2: Prompt Structure.} The structure the prompts use for the LLMs plays a crucial role in generating accurate answers. We observe that longer answers could adversely affect the evaluation metrics, especially when the actual answers are typically short. These changes aim at dealing with redundancies in LLM output. For example:\\
\fbox{%
\begin{minipage}{233pt}
\texttt{%
    \textbf{Question:} Who has joined the inquiry into Jackson's death? \\
    \textbf{Predicted Answer:} The Drug Enforcement Administration has joined the investigation into Jackson's death. \\
    \textbf{Actual Answer:} The Drug Enforcement Administration
}
\end{minipage}
}
To address this, we optimized the prompt structure utilizing prompt engineering enhancement methods\cite{white2023prompt}, incorporating more specific instructions such as 
\begin{quote}
    Please provide the answer in as few words as possible and do NOT repeat any word in the question, i.e. '\{question\}'.
\end{quote} 

This refinement enabled the model to produce answers more aligned with the actual responses, thereby improving model performances especially in precision.

\textbf{LLMs.} The choice of LLMs significantly influences the output quality. To overcome limitations observed with the GPT4ALL Falcon model, we experiment with different LLMs. Notably, the adoption of the Mistral 7B model, which is more responsive to instructions compared to the Falcon model, showed promising improvements\cite{jiang2023mistral}. This change in LLMs helped mitigate discrepancies in results attributable solely to the model's inherent structure.

\begin{comment}
After experimenting with different parameters focused on the comparison between cosine and pairwise distance functions, and the impact of varying the number of top sentences retrieved (TopN), the results indicate two key observations:
\begin{enumerate}
    \item Pairwise Distance Function Outperforms Cosine Similarity
    \begin{itemize}
        \item Effectiveness: The pairwise distance function demonstrated superior performance in terms of Exact Match (EM), Precision, Recall, and F1 Score compared to the cosine function. %For instance, with TopN set to 1, pairwise achieved an EM of 0.161435, Precision of 0.267, and F1 of 0.261, outperforming cosine on all these metrics.
        \item Efficiency: In terms of computational time, pairwise also exhibited a significant advantage. The time taken for pairwise with TopN set to 1 was  considerably less than the time taken for the cosine function under similar conditions. %(25480.5 seconds).
    \end{itemize}
    \item Optimal Configuration - Pairwise with Top 1\\
    The best results were observed when using the pairwise distance function with TopN set to 1. This configuration achieved the highest scores across all metrics while also being the most time-efficient. This suggests that focusing on the most relevant sentence, as determined by the pairwise distance function, yields the most accurate and efficient outcomes.
\end{enumerate}
\end{comment}

\subsection{NewsQA Experiments}
% TODO: Present the initial results before optimization.
% TODO: Discuss the shortcomings in these initial experiments. 

% Results on NewsQA based on different chunking and overlap using BERT embedding and Falcon model: 
% In establishing a baseline for our NewsQA experiment based on BERT embedding and Falcon model, 

Table~\ref{tab:baseline} shows baseline results based on segmentation and retrieval of sentences. Among these, the model achieved best performances (F1=0.261) with the pairwise distance function and top 1 hit for the QA context. The distance function favored longer sentences and it appears 1 sentence was about right, relatively speaking, for the model to identify correct answers. % The results, summarized in the table 1, provide a comprehensive overview of the system's performance across different configurations:

\begin{table}[htbp]
    \centering

    \setlength\tabcolsep{3pt}
    \begin{tabular}{lccccc}
        \toprule
        Distance & TopN & EM & Precision & Recall & F1 Score \\
        \midrule
        cosine & 1 & 0.128 & 0.210 & 0.244 & 0.210 \\ 
        cosine & 2 & 0.124 & 0.199 & 0.215 & 0.191 \\ 
        pairwise & 1 & 0.161 & 0.267 & 0.304 & 0.261 \\
        pairwise & 2 & 0.141 & 0.218 & 0.250 & 0.216 \\
        \bottomrule
    \end{tabular}
    \caption{NewsQA results with sentences for context}
    \label{tab:baseline}
\end{table}

From experiments using fixed chunks and overlaps, results revealed that a chunk size of 100 consistently produced the best results in terms of EM, Precision, Recall, and F1 Score, as shown in Table~\ref{tab:chunks}. Notably, the configuration with a chunk size of 100, using 0 overlap, cosine distance function and TopN set to 2, achieved the highest scores across these metrics. A chunk size of 100 strikes a balance between providing sufficient context for the model to follow and answer queries accurately and avoiding information overload with irrelevant data/context. Smaller chunks (e.g., 25 or 50) appear to lack comprehensive context, while larger ones (e.g., 200 or 300) potentially introduce noise and reduce precision. However, there are no clear relations demonstrated between the end result and distance function / TopN combinations.

\begin{table}[htbp]
    \centering
    \setlength\tabcolsep{3pt}
    \begin{tabular}{cccccccc}
    \toprule
     Chunk &  Overlap &     Dist &  TopN  &   EM &  Precision & Recall &   F1 \\
    \midrule
        25 &      0.0 & pairwise &     2  & 0.05 & 0.06 & 0.08 & 0.06 \\
        50 &      0.0 &   cosine &     4  & 0.06 & 0.06 & 0.09 & 0.06 \\
        50 &      0.0 &   cosine &     2  & 0.00 & 0.04 & 0.16 & 0.06 \\
       100 &      0.0 &   cosine &     1 5 & 0.25 & 0.39 & 0.39 & 0.36 \\
       \textbf{100} &  \textbf{0.0} &\textbf{cosine} & \textbf{2}  & \textbf{0.30} & \textbf{0.44} & \textbf{0.42} & \textbf{0.41} \\
       100 &      0.0 & pairwise &     2  & 0.28 & 0.42 & 0.42 & 0.39 \\
       100 &      0.1 & pairwise &     2  & 0.25 & 0.38 & 0.37 & 0.35 \\
       200 &      0.0 & pairwise &     1  & 0.17 & 0.24 & 0.25 & 0.23 \\
       200 &      0.2 &   cosine &     1  & 0.16 & 0.25 & 0.25 & 0.23 \\
       300 &      0.0 & pairwise &     1  & 0.14 & 0.23 & 0.23 & 0.21 \\
       300 &      0.0 &   cosine &     1  & 0.14 & 0.22 & 0.21 & 0.20 \\
    \bottomrule
    \end{tabular}
    \caption{NewsQA results with chunks for context}
    \label{tab:chunks}
\end{table}
%\vspace{-3mm}

\subsection{QAConv Experiments}
The baseline of the QAConv is based on the sentence segmentation of the story text and Table~\ref{tab:qaconv_baseline} is showing the evaluation results.

\begin{table}[htbp]
    \centering
    \setlength\tabcolsep{3pt}
    \begin{tabular}{cccc}
        \toprule
         EM & Precision & Recall & F1 Score \\
        \midrule
         0.040 & 0.127 & 0.246 & 0.147 \\ 
        \bottomrule
    \end{tabular}
    \caption{QAConv results with sentences for context}
    \label{tab:qaconv_baseline}
\end{table}
% Results on QAConv based on different chunking and overlap using BERT embedding and Falcon model: After applying the knowledge and methods we did research on, we managed to build a baseline for QAConv data to test its performance based on different chunk sizes and overlap percentages. The results in table 2 shows the system's general performance using different parameters combinations:

We experimented with similar combinations of chunking and overlapping on the QAConv data and obtained results in Table~\ref{tab:qaconv}, which indicates that the optimal performance, comparable to the NewsQA dataset, is achieved with a chunk size of 100 and an overlap percentage of 0. This combination strikes a balance between accurately retrieving relevant information and overall QA performance, as measured by the F1 score. While larger chunk sizes enhance the model's ability to identify and match relevant answers by providing more context, they also introduce a trade-off between having sufficient information vs. potential noise. Thus, finding the right balance between chunk size and overlap is crucial for maximizing the model's efficiency and accuracy.  %suggesting a need for strategic combinations of different chunk sizes to harness the benefits of both small and large chunks effectively.

% \vspace{-2mm}
\begin{table}[htbp]
    \centering
    \setlength\tabcolsep{3pt}
    \begin{tabular}{cccccc}
        \toprule
        Chunk & Overlap & EM & Precision & Recall & F1 Score \\ 
        \midrule
        400 & 0 & 0.061 & 0.127 & 0.263& 0.145 \\ 
        400 & 0.1 & 0.050 & 0.129 & 0.298 & 0.153 \\ 
        300 & 0 & 0.030 & 0.126 & 0.239 & 0.147 \\ 
        300 & 0.1 & 0.040 & 0.138 & 0.276 & 0.160 \\ 
        200 & 0 & 0.030 & 0.135 & 0.278 & 0.163 \\ 
        200 & 0.1 & 0.040 & 0.122 & 0.218 & 0.144 \\ 
        100 & 0 & 0.010 & 0.147 & 0.271 & 0.177 \\ 
        100 & 0.1 & 0.020 & 0.084 & 0.187 & 0.101 \\ 
        \bottomrule
    \end{tabular}
    \caption{QAConv result with chunks for context}
    \label{tab:qaconv}
\end{table}

\vspace{-4mm}

\section{Conclusion}
% TODO: Summarize the key findings and their implications.
% TODO: Suggest future directions for research in this area.

We introduce a novel framework to enhance performance in QA tasks through the integration of optimized vector retrieval and LLM instructions. Using retrieval augmentation, our approach encompasses document embedding, vector retrieval, and the assembly of context to improve QA generations. Our experiments explore various text segmentation techniques, e.g. sentence segments and fixed chunks, and similarity functions to test their impacts on QA performance. Findings reveal that the model based on a minimal chunk size of 100, with no overlap between chunks, delivers superior performance compared to models utilizing semantic segmentation at the sentence level. We provide examples of related QA tasks and shed light on the improvement of model performance within this two-stage framework. 

RAG-based QA models are increasingly popular because of the combined power of IR and LLMs without costly training or fine-tuning. By integrating optimized vector retrieval and instructions, our approach is focused on chunk-based retrieval strategies and fine-tuned prompt structures and has demonstrated improvements in QA precision. Our experiments on the two benchmark datasets show consistent results on the selection of chunk sizes. Retrieving larger chunks can potentially provide better “domain” knowledge for the GPTs while sacrificing some accuracy for the answers. While smaller chunks can provide a more targeted responses to questions, the granularity needs to be carefully measured to balance the scope for LLMs.
% The incorporation of diverse LLMs further enhances the system's effectiveness. 
The potential for applying these findings in fields requiring precise information retrieval is substantial. Future work could explore varying (dynamic) chunk sizes, strategies to retrieve and organize them, and their combined impacts on QA performances. % Additionally, adapting this approach to specialized domains presents an exciting avenue for practical applications. In essence, this research contributes a novel methodology to the evolving landscape of QA systems, paving the way for more accurate and efficient information retrieval solutions.

% 

% \begin{acks}
% % TODO: Acknowledge any assistance or contributions from others.

% \end{acks}

\clearpage
% TODO: List all the references in `references.bib` and use \cite{key} inline. 
\bibliographystyle{ACM-Reference-Format}
\bibliography{main}

% \appendix

\end{document}